\documentclass[twocolumn]{aastex61}

\usepackage{graphicx}	
\usepackage{amsmath}	
\usepackage{amssymb}	
\usepackage{bm}		
\usepackage{multirow}

\usepackage{placeins}

\received{xx}
\revised{xx}
\accepted{xx}
\submitjournal{ApJ}

\shorttitle{Different heating of the NS crust during multiple outbursts in MAXI J0556}
\shortauthors{Parikh et al.}

\begin{document}

\defcitealias{homan2014strongly}{Ho14}

\title{Different accretion heating of the neutron star crust during multiple outbursts in MAXI J0556$-$332}

\correspondingauthor{Aastha Parikh}
\email{a.s.parikh@uva.nl}

\author{Aastha S. Parikh}
\affil{Anton Pannekoek Institute for Astronomy, University of Amsterdam, Postbus 94249, NL-1090 GE Amsterdam, the Netherlands}

\author{Jeroen Homan}
\affiliation{SRON, Netherlands Institute for Space Research, Sorbonnelaan 2, 3584 CA Utrecht, The Netherlands}
\affil{Eureka Scientific, Inc., 2452 Delmer Street, Oakland, California 94602}

\author{Rudy Wijnands}
\affiliation{Anton Pannekoek Institute for Astronomy, University of Amsterdam, Postbus 94249, NL-1090 GE Amsterdam, the Netherlands}

\author{Laura Ootes}
\affiliation{Anton Pannekoek Institute for Astronomy, University of Amsterdam, Postbus 94249, NL-1090 GE Amsterdam, the Netherlands}

\author{Dany Page}
\affiliation{Instituto de Astronom\'{i}a, Universidad Nacional Aut\'{o}noma de M\'{e}xico, Mexico D.F. 04510, Mexico}

\author{Diego Altamirano}
\affiliation{Department of Physics and Astronomy, Southampton University, Southampton SO17 1BJ, UK}

\author{Nathalie Degenaar}
\affiliation{Anton Pannekoek Institute for Astronomy, University of Amsterdam, Postbus 94249, NL-1090 GE Amsterdam, the Netherlands}

\author{Edward F. Brown}
\affiliation{Department of Physics and Astronomy, Michigan State University, 567 Wilson Rd, East Lansing, MI 48864, USA}

\author{Edward Cackett}
\affiliation{Department of Physics \& Astronomy, Wayne State University, 666 W. Hancock St, Detroit, MI 48201, USA}

\author{Andrew Cumming}
\affiliation{Department of Physics and McGill Space Institute, McGill University, 3600 Rue University, Montreal QC, Canada H3A 2T8}

\author{Alex Deibel}
\affiliation{Astronomy Department, Indiana University, Bloomington, IN 47405}

\author{Joel K. Fridriksson}
\affiliation{Anton Pannekoek Institute for Astronomy, University of Amsterdam, Postbus 94249, NL-1090 GE Amsterdam, the Netherlands}

\author{Dacheng Lin}
\affiliation{Space Science Center, University of New Hampshire, Durham, NH 03824, USA}

\author{Manuel Linares}
\affiliation{Departament de F{\'i}sica, EEBE, Universitat Polit{\`e}cnica de Catalunya, c/ Eduard Maristany 10, 08019 Barcelona, Spain}

\author{Jon M. Miller}
\affiliation{Department of Astronomy, University of Michigan, 500 Church Street, Ann Arbor, MI 48109, USA}



\begin{abstract}

The transient neutron star (NS) low-mass X-ray binary MAXI J0556$-$332 provides a rare opportunity to study NS crust heating and subsequent cooling for multiple outbursts of the same source. We examine {\it MAXI}, {\it Swift}, {\it Chandra}, and {\it XMM-Newton} data of MAXI J0556$-$332 obtained during and after three accretion outbursts of different durations and brightness. We report on new data obtained after outburst III. The source has been tracked up to $\sim$1800 d after the end of outburst I. Outburst I heated the crust strongly, but no significant reheating was observed during outburst II. Cooling from $\sim$333 eV to $\sim$146 eV was observed during the first $\sim$1200 d. Outburst III reheated the crust up to $\sim$167 eV, after which the crust cooled again to $\sim$131 eV in $\sim$350 d. We model the thermal evolution of the crust and find that this source required a different strength and depth of shallow heating during each of the three outbursts. The shallow heating released during outburst I was $\sim$17 MeV nucleon$^{-1}$ and outburst III required $\sim$0.3 MeV nucleon$^{-1}$. These cooling observations could not be explained without shallow heating. The shallow heating for outburst II was not well constrained and could vary from $\sim$0--2.2 MeV nucleon$^{-1}$, i.e., this outburst could in principle be explained without invoking shallow heating. We discuss the nature of the shallow heating and why it may occur at different strengths and depths during different outbursts. 
\end{abstract}




\section{Introduction} \label{sec:intro}

Transient neutron stars (NSs) in low-mass X-ray binaries (LMXBs) are excellent laboratories to study dense matter physics. These systems experience accretion outbursts separated by periods of quiescence. 
During outbursts, the accreted matter compresses the NS surface, inducing heat-releasing nuclear reactions deep in the crust \citep{haensel1990non,haensel2008models,steiner2012deep} that disrupt the crust-core thermal equilibrium. Once the source transitions into quiescence the crust cools to restore equilibrium with the core. Tracking this crustal cooling and fitting theoretical models to study its evolution allows us to infer NS crust properties \citep[e.g.,][]{brown2009mapping}. So far, eight NS LMXBs that show crustal cooling have been studied \citep[see][for a review]{wijnands2017review}. In addition to the standard deep crustal heating ($\rho \sim10^{12}$--$10^{13}$ g cm$^{-3}$), a shallow heat source ($\rho \sim10^{8}$--$10^{10}$ g cm$^{-3}$) is required to explain the observed cooling curve of many of these sources. The origin of this shallow heat source is unknown and is important to resolve because the cooling curve can be used to constrain a number of different aspects of crust physics \citep[such as conductivity of the crust and pasta, and the core specific heat;][]{brown2009mapping,horowitz2015disordered,cumming2017lower}. Most systems need $\sim$1--2 MeV nucleon$^{-1}$ of shallow heating to explain their cooling curves (e.g., \citeauthor{degenaar2014probing} \citeyear{degenaar2014probing}, \citeauthor{parikh2017potential} \citeyear{parikh2017potential}; \citeauthor{wijnands2017review} \citeyear{wijnands2017review}).

\begin{figure*}
\centering
\includegraphics[scale=0.745]{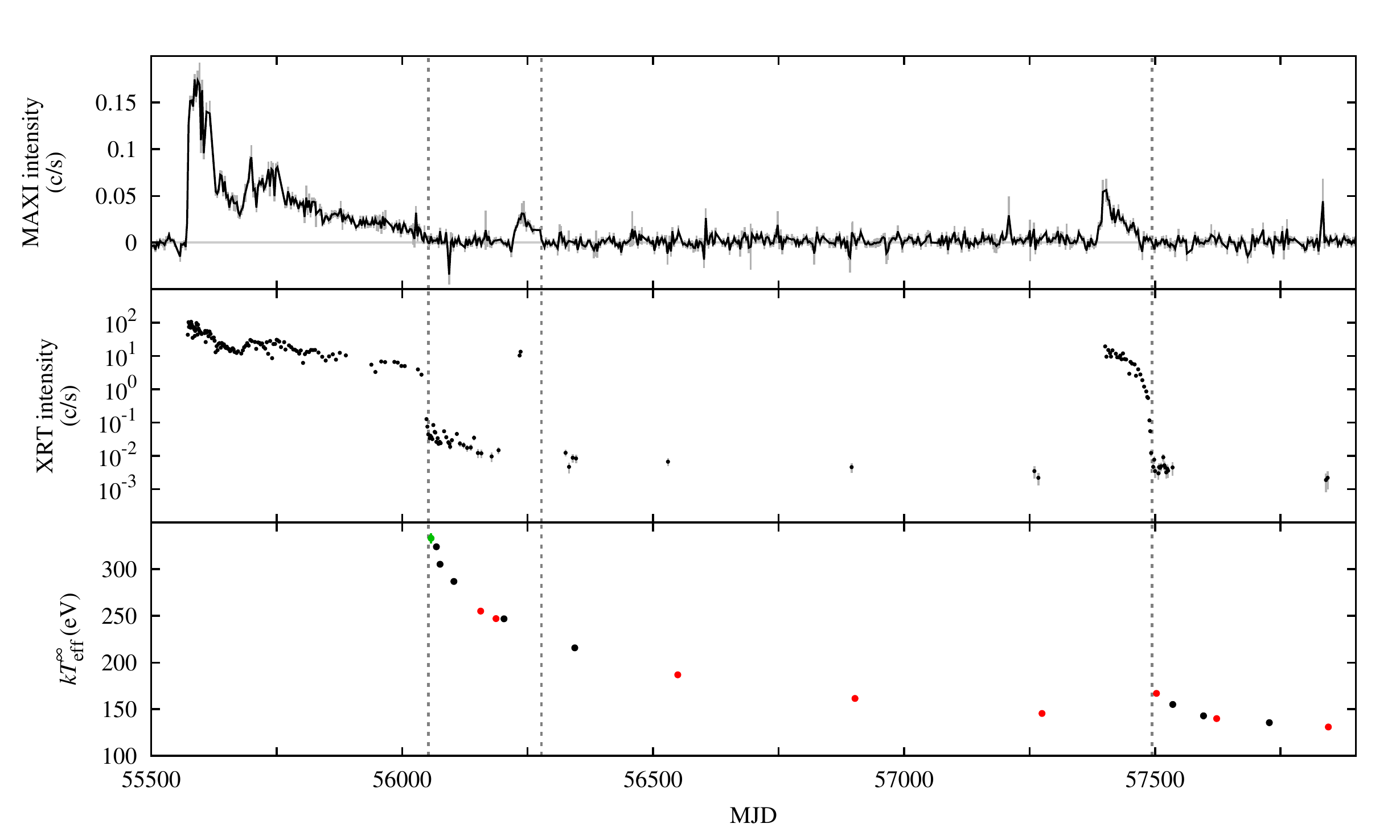}
\caption{Long-term light curves of J0556 from {\it MAXI}/GSC (top panel, 2--4 keV) and {\it Swift}/XRT (middle panel, 0.5--10 keV). In the bottom panel we show the temperature evolution of the source during its quiescent periods (the green, black, and red points indicate the {\it Swift}/XRT, {\it Chandra}, and {\it XMM-Newton} data, respectively). The dotted grey lines indicate the time of transition to quiescence after each outburst.}
\label{fig_lc}
\end{figure*}

The transient NS LMXB MAXI J0556$-$332 (hereafter J0556) was discovered on 2011 January 11 \citep{matsumura2011maxi} and exhibited a $\sim$16 month outburst. The source showed a second outburst in 2012 that lasted $\sim$2 months \citep{sugizaki2012re} 
and a third $\sim$3 month outburst in 2016 \citep{negoro2016maxi}. In Figure \ref{fig_lc} (top panel) we show the {\it MAXI} light curve with all three outbursts. \citet{sugizaki2013spectral} examined the spectral data obtained using the {\it MAXI}, {\it Swift}, and {\it RXTE} when the source was in outburst, and constrained the source distance to be $>$17 kpc. \citet[][hereafter \citetalias{homan2014strongly}]{homan2014strongly} studied the {\it Swift}, {\it Chandra}, and {\it XMM-Newton} spectra of the source after its outbursts  and found a distance of $\sim$45 kpc. Such a large distance was further supported by the  46$\pm$15 kpc distance the same authors obtained when comparing the X-ray colour-colour and hardness-intensity diagram of J0556 with those observed for other, similarly bright ($\sim$Eddington limited) NS LMXBs (i.e., the so-called Z sources).

\citetalias{homan2014strongly} studied J0556 in quiescence after outburst I and II and found it to have a very hot NS crust. They showed that outburst II did not seem to reheat the NS crust. \citet{deibel2015strong} showed that J0556 released a very large amount of shallow heating during outburst I of $\sim$10 MeV nucleon$^{-1}$ to explain its crust cooling evolution, the largest required by any NS LMXB source studied so far. 
During outburst II, the shallow heating mechanism was inactive or at a much reduced level compared to outburst I. Here we present cooling observations of J0556 after outburst III demonstrating that the source was reheated during this outburst, requiring a small but significant amount of shallow heating. Therefore, the shallow heating mechanism is not just active or inactive but can indeed be active at different strengths.

\section{Observations, Data Analysis, and Results}
J0556 has been observed in quiescence by the {\it Swift}, {\it Chandra}, and {\it XMM-Newton} observatories. We use {\it MAXI} and the X-ray Telescope \citep[XRT;][]{burrows2005swift} aboard {\it Swift} to track the variability of the source during its outbursts. We report on eight new {\it Chandra} and {\it XMM-Newton} observations of this source. For uniformity, we also reanalyse all observations reported by \citetalias{homan2014strongly}. Table \ref{tab_log} shows the log of quiescent observations.

\subsection{\textbf{\textit{MAXI}}}
The outburst evolution of J0556 was observed using the {\it MAXI}/Gas Slit Camera \citep[GSC;][]{mihara2011gas}. We downloaded the light curve from the {\it MAXI} archive\footnote{http://maxi.riken.jp/top/slist.html} for the 2--4 keV range as \citetalias{homan2014strongly} suggest that this energy range results in the highest signal-to-noise ratio. Similar to their analysis, we remove data that have error bars larger 0.025 counts s$^{-1}$ and apply a 3 day rebinning. 

\subsection{\textbf{\textit{Swift}/XRT}}
\label{sect_swift_int}
{\it Swift}/XRT was also used to monitor the evolution of J0556. 
The raw data were processed with \texttt{HEASOFT} (version 6.17) using \texttt{xrtpipeline}. The light curve and spectra were extracted using \texttt{XS\textsc{elect}} (version 2.4c). A circular source extraction region with a radius of 40$''$ was used. For the background extraction region we used an annulus of inner and outer radii 50$''$ and 80$''$, respectively. 

Four observations after outburst I were combined into one interval to obtain constraints on the earliest crust cooling phase (observation ID [obsID]: 00032452004--00032452007; from 2012 May 7 to 2012 May 11). The Photon Counting mode data from these observations were stacked into a single event file and the spectrum was extracted using the same source and background regions as those used for the light curve. The ancillary response file was generated using \texttt{xrtmkarf} and the appropriate response matrix file `swxpc0to12s6$\_$20110101v014.rmf' was used. The 0.3--10 keV spectrum was binned to have a minimum of 5 counts per bin using \texttt{grppha}.

\subsection{\textbf{\textit{Chandra}}}
J0556 was observed ten times with the {\it Chandra}/ACIS-S \citep{garmire2003advanced} in the \texttt{FAINT} mode using the 1/8 sub-array. 
Two observations \citepalias[ObsID: 14429 and 14227; see Table 1 of][]{homan2014strongly} took place during episodes of temporary increases in accretion rate during the first $\sim$100 d after the end of outburst I. Since we are only interested in the crust cooling behaviour we do not discuss these data \citepalias[see][for more information about these observations]{homan2014strongly}. \texttt{CIAO} (version 4.9) was used to process the raw data of the remaining eight observations. We examined the source light curves for possible episodes of background flaring, but none were found. We used circular source extraction regions with a radius of 2$''$--3$''$ (depending on source brightness). The background extraction region used was an annulus with inner and outer radii 10$''$ and 20$''$, respectively. The spectra were extracted using \texttt{specextract}. The point-source aperture-corrected \texttt{arf} files, as generated by \texttt{specextract}, were used. The 0.3--10 keV spectra were grouped using \texttt{dmgroup} to have a signal-to-noise ratio of 4.5.

\subsection{\textbf{\textit{XMM-Newton}}}
{\it XMM-Newton} \citep{struder2001european} was used to observe J0556 eight times using all three European Photo Imaging Cameras (EPIC) -- MOS1, MOS2, and pn.  
The raw data were reduced using the Science Analysis System (\texttt{SAS}; version 16.0) and processed using \texttt{emproc} and \texttt{epproc} for the MOS and pn detectors, respectively. The observations were checked for background flaring by examining the light curves for the $>$10 keV range for the MOS detectors and 10--12 keV range for the pn detector. Depending on the average source brightness during an observation, we removed data of $>$0.08--0.16 counts s$^{-1}$ and data of $>$0.32--0.4 counts s$^{-1}$ for the MOS and pn detectors, respectively. Circular source and background extraction regions were used for the spectral extraction. The optimal source region to be used was determined with \texttt{eregionanalyse}. Depending on the source brightness, regions of radii 19$''$--45$''$ were used for the MOS detectors and regions of radii 19$''$--37$''$ were used for the pn detector. A background region of radius 50$''$ was used for each observation, placed on the same CCD that the source was located on. The position of the background region as suggested by the \texttt{ebkgreg} tool was used. 
The redistribution matrix file and ancillary response function were generated using \texttt{rmfgen} and \texttt{arfgen}. The 0.3--10 keV spectra were binned using \texttt{specgroup}, with the signal-to-noise ratio set to 4.5.

\label{sec_res}
\subsection{Spectral Fitting}
\label{sec_res_spec}

All {\it Chandra} and {\it XMM-Newton} 0.3--10 keV spectra were fit simultaneously in \texttt{XS\textsc{pec}} (version 12.9; \citeauthor{arnaud1996xspec} \citeyear{arnaud1996xspec}) using $\chi^2$ statistics. We used the NS atmosphere model \texttt{nsa} \citep{zavlin1996model}. The \texttt{nsatmos} model cannot be used as the source was very hot at the start of the cooling phase \citepalias[see][for details]{homan2014strongly}. The temperature was left free for each individual {\it XMM-Newton} and {\it Chandra} observation. However, this parameter was tied between the MOS and pn cameras for a given {\it XMM-Newton} observation. All the temperatures were converted into the effective temperature measured by an observer at infinity. The mass and radius of the NS were fixed to 1.4 $M_\odot$ and 10 km \citep[since in the \texttt{nsa} model the surface gravity has only been calculated for this combination and does not give accurate results for other mass and radius values;][]{zavlin1996model,heinke2006hydrogen}. The magnetic field parameter was set to zero (indicating a non-magnetised star). The normalization (1/$D^2$; {\it D} is the distance) and the $N_\text{H}$ were free parameters but were tied between all observations. The $N_\text{H}$ was modeled using \texttt{tbnew}$\_$\texttt{feo}\footnote{http://pulsar.sternwarte.uni-erlangen.de/wilms/research/tbabs/} with \texttt{WILM} abundances \citep{wilms2000absorption} and \texttt{VERN} cross-sections \citep{verner1996atomic}. The Oxygen and Iron abundances for \texttt{tbnew}$\_$\texttt{feo} were fixed to 1.

We use the \texttt{pileup} component to model the pile-up in the {\it Chandra} data\footnote{We use values suggested by the {\it Chandra} pile-up guide: http://cxc.harvard.edu/ciao/download/doc/pileup\_abc.pdf}. We set the maximum number of photons considered for pile up in a single frame to 3, the grade correction for single photon detection to 1, the PSF fraction is fixed to 0.95, the number of regions and the \texttt{FRACEXPO} keyword are both set to 1. We find that the fits are not very sensitive to the alpha parameter (see also \citetalias{homan2014strongly}) and we fix this value to 0.6. The frame time, an input parameter for \texttt{pileup}, was set to 0.4 s for the {\it Chandra} observations. To prevent the pile-up model from affecting the {\it XMM-Newton} data we set the frame time for these observations to a very small value ($10^{-6}$ s). We also use the multiplicative model \texttt{constant} to allow for normalisation offsets between the {\it Chandra} and {\it XMM-Newton} observatories. For the {\it Chandra} data we used a constant of 1 and for the various {\it XMM-Newton} detectors -- MOS1, MOS2, and pn the constants were calculated from Table 5 of \citet[][$C_\text{MOS1} = 0.983$, $C_\text{MOS2} = 1$, $C_\text{pn} = 0.904$]{plucinsky2017SNR}.


\begin{table*}
\centering
\caption{Log of the quiescent observations.$^\text{a}$}
\label{tab_log}
\begin{tabular}{lllccccc}
\hline
& Instrument & ObsID & Days since &Exposure		& $kT_\text{eff}^\infty$ & $F_\text{X}$& $L_\text{X}$ \tabularnewline
& 					&			&end of&time$^{\text{b}}$ & (eV)							& ($ \times 10^{-14}$ & ($\times 10^{33}$ \tabularnewline
& 					&		&outburst I&(ksec)	& 						& erg cm$^{-2}$ s$^{-1}$)& erg s$^{-1}$)\tabularnewline
\hline
\multicolumn{8}{>{\centering\arraybackslash}p{16cm}}{{\it After Outburst I}} \tabularnewline

1 & {\it Swift} & Interval 1$^\text{c}$ & 		5.4 	 	&	10.1 & 				$333.1\pm5.5   	$	&	$109.7 \pm	 6.6 $ &   			$249.5\pm		 15.1$\tabularnewline
2 & {\it Chandra} & 14225 & 				16.0 		&9.2 &					$324.1\pm2.5$		&	$106.3  \pm	 3.6 $ &			$241.9\pm	  	  8.3$\tabularnewline
3 & {\it Chandra} & 14226 & 				23.3 		&9.1 &					$305.4\pm2.5$		&	$79.7   \pm	 2.8 $ &			$181.2\pm	  	  6.3$\tabularnewline
4 & {\it Chandra} & 14433 & 				51.0 		&18.2 &					$286.9\pm1.9	$	&	$62.4   \pm	 1.8 $ &			$142.0\pm	  	  4.0$\tabularnewline
5 & {\it XMM-Newton} & 0700380901 & 			104.3 		 &28.2, 27.6, 21.5  &			$255.2\pm0.9	$	&	$37.9   \pm	 0.6 $ &			$86.2 \pm	 	  1.3$\tabularnewline
6 & {\it XMM-Newton} & 0700381201 & 			134.9 		 &24.4, 23.7, 17.5 &			$247.1\pm1.0	$	&	$32.6   \pm	 0.7 $ &			$74.1 \pm	 	  1.4$\tabularnewline
7 & {\it Chandra} & 14434 & 				150.8 		&18.2 &					$246.8\pm2.1	$	&	$33.5   \pm	 1.3 $ &			$76.2 \pm	 	  2.8$\tabularnewline
\multicolumn{8}{>{\centering\arraybackslash}p{16cm}}{{\it After Outburst II}} \tabularnewline

8 & {\it Chandra} & 14228 & 				291.7 		&22.7 &					$215.8\pm2.2	$	&	$19.1  \pm	 0.8 $ &			$43.5 \pm	 	 1.9 $\tabularnewline
9 & {\it XMM-Newton} & 0725220201 & 			496.8 		 &44.3, 42.3, 34.9 &			$186.8\pm1.0	$	&	$10.2  \pm	 0.2 $ &			$23.2 \pm	 	 0.6 $\tabularnewline
10 & {\it XMM-Newton} & 0744870201 & 			850.1 		 &77.3, 78.6, 59.9 &			$161.5\pm0.8	$	&	$5.5  \pm	 0.1 $ &			$12.6 \pm	 	 0.3 $\tabularnewline
11 & {\it XMM-Newton} & 0762750201 & 			1222.6 		 &83.9, 72.3, 57.5	&		$145.4\pm1.0	$	&	$3.5  \pm	 0.1 $ &			$7.9 \pm	 	 0.3 $\tabularnewline
\multicolumn{8}{>{\centering\arraybackslash}p{16cm}}{{\it After Outburst III}} \tabularnewline	
	
12 & {\it XMM-Newton} & 0784390301 & 			1450.7 		 &20.4, 18.6, 18.6 &			$166.9\pm1.8	$	&	$6.9  \pm	 0.5 $ &			$15.7 \pm	 	 1.1 $\tabularnewline
13 & {\it Chandra} & 18335 & 				1483.2 		&27.3 &					$155.1\pm2.8	$	&	$4.8  \pm	 0.4 $ &			$10.9\pm	 	 1.0 $\tabularnewline
14 & {\it Chandra} & 18336 & 				1544.3 		&27.1 &					$142.9\pm3.3	$	&	$3.4  \pm	 0.5 $ &			$7.8 \pm	 	 1.0 $\tabularnewline
15 & {\it XMM-Newton} & 0784390401 & 			1570.6 		 &45.9, 41.3, 39.5 &			$139.9\pm1.4	$	&	$3.0  \pm	 0.1 $ &			$6.7 \pm	 	 0.3 $\tabularnewline
16 & {\it Chandra} & 18337 & 				1675.5 		&27.3 &					$135.6\pm3.5	$	&	$2.6  \pm	 0.3 $ &			$6.0 \pm	 	 0.7 $\tabularnewline
17 & {\it XMM-Newton}& 0782670201 &			1793.1 		  &61.0, 58.3, 36.7 &			$130.9\pm1.5	$	&	$2.3  \pm	 0.1 $ &			$5.1 \pm	 	 0.3 $\tabularnewline

\hline
\multicolumn{8}{p{16cm}}{\textsuperscript{$^\text{a}$}\scriptsize{All errors are 1$\sigma$. The distance and $N_\mathrm{H}$ were fixed to 43.6 kpc and $3.2 \times 10^{20}$ cm$^{-2}$. The unabsorbed fluxes and luminosities are quoted for 0.5--10 keV.}} \tabularnewline
\multicolumn{8}{p{16cm}}{\textsuperscript{$^\text{b}$}\scriptsize{The {\it XMM-Newton} effective exposure times have been displayed as `MOS1, MOS2, pn'.}}\tabularnewline
\multicolumn{8}{p{16cm}}{\textsuperscript{$^\text{c}$}\scriptsize{See Section \ref{sect_swift_int}.}}\tabularnewline
\end{tabular}
\end{table*}

The simultaneous fit of all the {\it Chandra} and {\it XMM-Newton} data resulted in a column density of $N_\text{H} = (3.2 \pm 0.5) \times 10^{20}$ cm$^{-2}$ and a distance of $43.6_{-1.6}^{+0.9}$ kpc. While calculating the error on the temperature we fixed the best fit $N_\text{H}$ and distance values as they are not expected to vary between observations \citep[see also, e.g.,][\citetalias{homan2014strongly}]{wijnands2004monitoring}. 
The parameters from the obtained fit ($\chi^2_\nu/\text{d.o.f.}$ = 1.02/715) are shown in Table \ref{tab_log}. Changing the mass, radius, $N_\mathrm{H}$, and distance will change the absolute $kT_\text{eff}^\infty$ but the trend, which helps us understand crust physics, will not change significantly \citep[e.g.,][]{cackett2008cooling}. The brightest {\it Chandra} observations had a pile-up fraction of $\sim$2 per cent.

Due to the small number of counts per bin in the {\it Swift}/XRT spectra (0.3--10 keV), they were fit separately with W-statistics (background subtracted Cash statistics) using the same $N_\text{H}$ and distance as used for the {\it Chandra} and {\it XMM-Newton} spectral fitting. We allow for a normalisation offset of $C_\text{XRT} = 0.872$ \citep{plucinsky2017SNR}.


Figure \ref{fig_lc} shows the {\it MAXI} (top panel) and {\it Swift}/XRT (middle panel) light curves of J0556. The bottom panel of Figure \ref{fig_lc} shows the cooling evolution of J0556, suggesting a strong decrease in the $kT_\text{eff}^\infty$ after the end of outburst I. Outburst II did not reheat the crust significantly and the cooling appeared to continue along its previous trend \citepalias[see also][]{homan2014strongly}. Overall, the crust cooled from $\sim$333 eV to $\sim$146 eV, after the outburst I and II. Outburst III caused the crust to be reheated significantly (to $\sim$167 eV) and was followed by subsequent cooling (to $\sim$131 eV approximately 350 days after the end of this recent outburst). 

\subsection{Modelling the quiescent thermal evolution}
\label{sec_nscool}
We model the outbursts and quiescence of J0556 using the crustal heating/cooling code \texttt{NSC\textsc{ool}} \citep{page2013forecasting,page2016nscool}. We account for the accretion rate variability during the outbursts and model all three outbursts collectively \citep[using the methods of][]{ootes2016,ootes2017}. The transition to quiescence after outburst I occurred on MJD 56052.1 \citepalias[see][]{homan2014strongly}. We use the methodology described by \citet{fridriksson2010rapid} to calculate the time of transition to quiescence by fitting an exponential to the rapidly decaying trend at the end of the outburst and a straight line to the quiescent points soon after the end of the outburst. We used the {\it MAXI} and {\it Swift}/XRT light curve, respectively, for this calculation after outburst I and II and determined the transition time to be MJD 56277.4 and MJD 57494.2.

We account for accretion rate variability during the outbursts by tracking the daily average rate, which is determined from the daily averaged {\it MAXI} and {\it Swift}/XRT count rates. If data from both instruments are available on the same day, the {\it Swift}/XRT data are used. \citet{sugizaki2013spectral} report the bolometric flux ($F_\mathrm{bol}$) from J0556 at six different instances during the first outburst (no such reports are available for the other two outbursts). We use these values to calculate a count rate to $F_\mathrm{bol}$ conversion constant for the {\it MAXI} and {\it Swift}/XRT data using count rates from the same days as the six reported $F_\mathrm{bol}$ values. This constant was calculated for each of the six observations for both instruments and the final constant used for each instrument was the averaged value. The obtained constants are $C_\mathrm{\it{MAXI}}$ = $2.353 \times 10^{-8}$ erg cm$^{-2}$ count$^{-1}$ and $C_\mathrm{\it{Swift}}$ = $4.957 \times 10^{-11}$ erg cm$^{-2}$ count$^{-1}$. The individual count rate to $F_\mathrm{bol}$ conversion constants differ at most by a factor of $\sim$2 from each other. We tested our \texttt{NSC\textsc{ool}} model results by using these minimum or maximum calculated individual conversion constant values (instead of the averaged ones) for all three outbursts as well as by using different values for different outbursts. We find that this only inconsequentially changes our inferred NS parameters without affecting our main conclusions. The $F_\mathrm{bol}$ was used to calculate the daily average accretion rate using $$\dot{M} = \frac{F_{\mathrm{bol}} 4 \pi D^2}  {\eta\, c^2}$$ where $\eta$ (= 0.2) indicated the efficiency factor and $c$ is the speed of light. Using this we obtained the fluences of the three outbursts to be $\sim$4.3 $\times$ 10$^{-2}$ erg cm$^{-2}$, $\sim$2.1 $\times$ 10$^{-3}$ erg cm$^{-2}$, $\sim$4.4 $\times$ 10$^{-3}$ erg cm$^{-2}$, respectively.

The NS mass and radius used for the \texttt{NSC\textsc{ool}} models are the same as those used for the spectral fits. The distance used is that calculated from the best fit to the spectra (Section \ref{sec_res_spec}). The \texttt{NSC\textsc{ool}} adjustable parameters are the impurity factor in the crust ($Q_\mathrm{imp}$), the column depth of light elements in the envelope ($y_\mathrm{light}$), the core temperature prior to the first outburst ($T_0$), and the strength ($Q_\mathrm{sh}$) and depth ($\rho_\mathrm{sh}$) of the shallow heating. We model the shallow heating with different parameters during each outburst. Furthermore, we also allow the envelope composition of the crust to vary for each outburst \citep[][]{ootes2017}. The best fit is found using a $\chi^2$ minimisation algorithm and  all errors are calculated for the 1$\sigma$ confidence range. 

\begin{figure}
\centering
\includegraphics[scale=0.46]{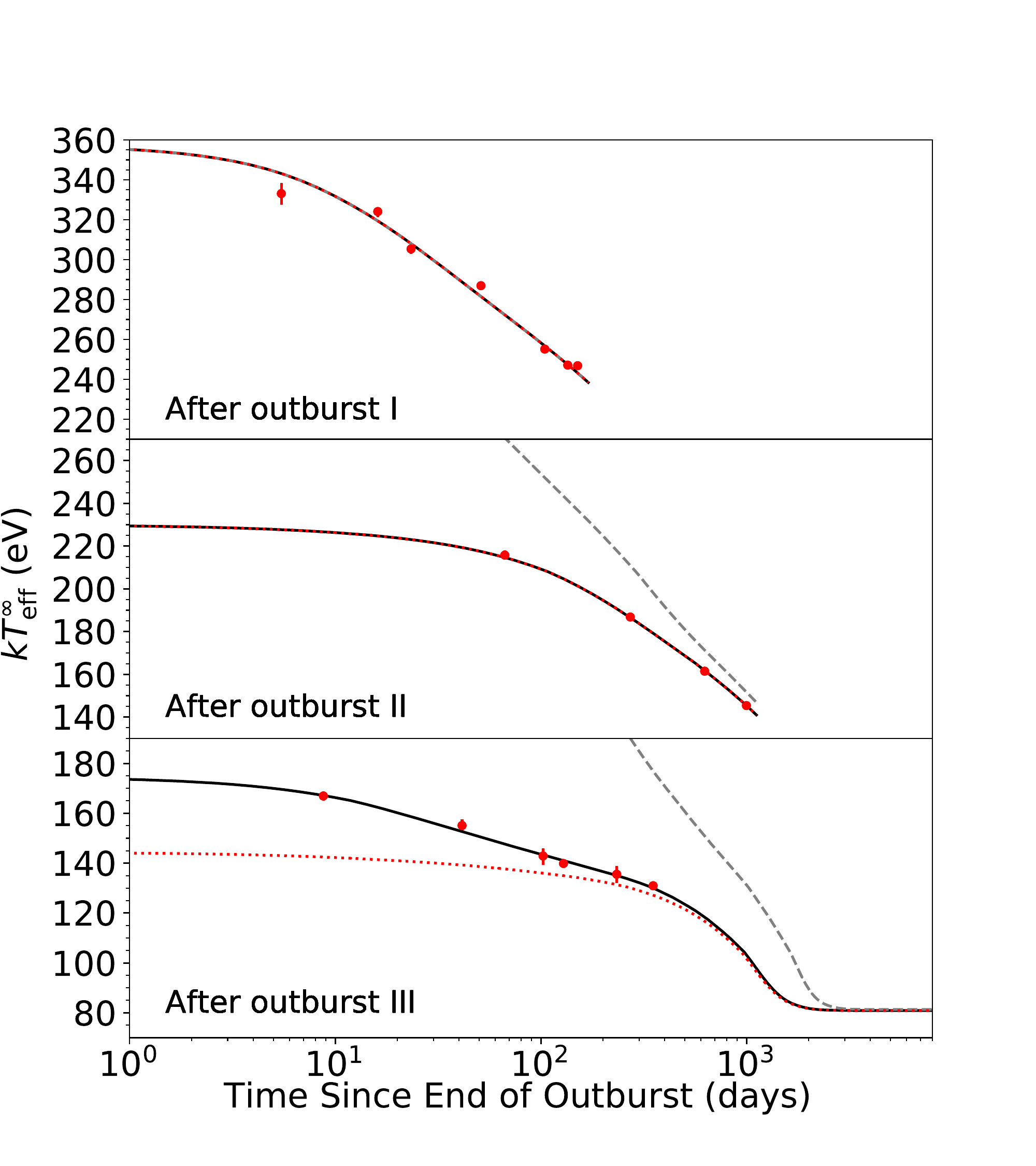}
\caption{The $kT_\text{eff}^\infty$ evolution of J0556 in quiescence with the cooling curves calculated using \texttt{NSC\textsc{ool}}. The solid black line shows the best fit model with all three outbursts having different shallow heating parameters. The dashed grey line shows the outcome if the same shallow heating parameters as inferred for outburst I were used for all three outbursts. The dotted red line shows the model when no shallow heating was assumed to be active during outburst II and III.}
\label{fig_cooling}
\end{figure}

\begin{table}
\centering
\caption{Shallow heating parameters from \texttt{NSC\textsc{ool}}.}
\label{tab_fit_NSCool}
\begin{tabular}{lcc}
\hline
Outburst& $Q_\mathrm{sh}$& $\rho_\mathrm{sh}$\tabularnewline
& (MeV nucleon$^{-1}$)&($\times 10^{9}$ g cm$^{-3}$ )\tabularnewline
\hline
I	& $17.0_{-0.7}^{+2.2}$	&	$5.3_{-0.5}^{+0.2}$ \tabularnewline
\multirow{2}*{II$^\text{a}$}	& 0&	--  \tabularnewline
& $2.2 \pm 0.7$ &	$33.5 \pm 0.8$  \tabularnewline
III	& $0.33 \pm 0.03$	&	$1.6 \pm 1.3$ \tabularnewline
\hline
\multicolumn{3}{p{7.5cm}}{\textsuperscript{$^\text{a}$}\scriptsize{Note that the parameters for outburst II are given for a range of values.}} \tabularnewline
\end{tabular}
\end{table}

Our modelling shows that heating only by deep crustal reactions (in combination with changes in the envelope composition) cannot explain the observations of J0556. A significant amount of shallow heating is necessary to explain the data. 
 $Q_\mathrm{sh}$ $\sim$17 MeV nucleon$^{-1}$ is needed to explain the high temperatures after outburst I. The best fit model ($\chi^2_\nu/\mathrm{d.o.f.} = 2.8/9$) is shown by the solid black line in Figure \ref{fig_cooling} and the parameters are listed in Table \ref{tab_fit_NSCool}. Outburst II and III cannot be explained with the same shallow heating as required for outburst I, not even if the envelope composition is allowed to change \citep{brown2002variability}. The dotted gray line in Figure \ref{fig_cooling} shows a model that assumes that during all three outbursts the shallow heating mechanism released $Q_\mathrm{sh}$ $\sim$17 MeV nucleon$^{-1}$, clearly showing that this would result in much hotter crusts than what we have observed.  Outburst III could be modelled with $Q_\mathrm{sh}$ $\sim$0.3 MeV nucleon$^{-1}$ at shallower depths than that required for outburst I. Outburst III could not be modelled without shallow heating. This is shown by the dotted red line in Figure \ref{fig_cooling} which indicates a model for which we assumed that the shallow heating mechanism was not active during outburst II and III. This model shows that outburst II can be explained without shallow heating. However, the uncertainties on the shallow heating parameters after outburst II are large, ranging from (a) no shallow heating ($Q_\mathrm{sh}$ $\sim$0 MeV nucleon$^{-1}$) when this heating source is at relatively shallow depths (i.e. the $\rho_\mathrm{sh}$ corresponding to outburst I or outburst III) as well as (b) a large $Q_\mathrm{sh} \sim$2.2 MeV nucleon$^{-1}$ very deep in the crust (at $\rho_\mathrm{sh}$ $\sim$ 33.5$\times 10^{9}$ g cm$^{-3}$). This is because the $Q_\mathrm{sh}$ and $\rho_\mathrm{sh}$ are correlated and the modelled curve from these two possibilities have a very similar shape.

In addition to constraining the shallow heating parameters, we found a core temperature of $T_0$ = ($5.5\pm0.4$)$\times 10^7$ K. Assuming a $Q_\mathrm{imp}$ = 1 throughout the crust fit the data well. Increasing the $Q_\mathrm{imp}$, to $\sim$20 as suggested by \citet{horowitz2015disordered}, in the pasta layer (which extends from $\rho \sim$10$^{13}$ g cm$^{-3}$ to the crust-core boundary) reduced the fit quality, indicating that the crust of this source has a high thermal conductivity throughout. 

Our best fit model indicates that a relatively light envelope is necessary after outburst I ($y_\mathrm{light}$ $\sim$3.1 $\times 10^{9}$ g cm$^{-2}$). 
The data after outburst I cannot be adequately fit using a heavy element envelope as the light elements are necessary to raise the effective temperature to that we observe. The envelope composition systematically raises or drops the observed trend whereas a change in the shallow heating changes the slope of the trend itself, therefore they are only partially degenerate in determining the cooling trend. The best fit indicates that a similar envelope composition ([2.1 -- 3.1]$\times 10^{9}$ g cm$^{-2}$) is present after all three outbursts. Further extensive investigation into the \texttt{NSC\textsc{ool}} parameters is not the aim of our paper. 

\section{Discussion}

We studied the NS LMXB J0556 in quiescence. Since the NS mass, radius, and distance remain the same between outbursts, studying the source after multiple outbursts allows us to investigate which parameters affecting the heating and cooling behaviour of the crust may change. As reported by \citetalias{homan2014strongly}, we found a strongly heated crust after outburst I. The amount of shallow heating during outburst I was very large, at $Q_\mathrm{sh}$ $\sim$17 MeV nucleon$^{-1}$. During outburst II the shallow heating mechanism may have been inactive \citep[see also][for similar results of outburst I and II]{deibel2015strong}. However, it should be noted that the first pointing was obtained long after the end of outburst II ($\sim$70 d) and $Q_\mathrm{sh}$ could not be constrained well. Up to $\sim$2.2 MeV nucleon$^{-1}$ was still allowed during this outburst if the heating occurred at much larger depths. We present new quiescent observations after the end of outburst III. Shallow heating (at $\sim$0.3 MeV nucleon$^{-1}$) needs to have been active during outburst III to explain the reheating we observed after the end of the outburst since the deep crustal reactions alone cannot account for this reheating. Based on the well constrained shallow heating parameters found after outburst I and III we find that this heating mechanism may release different amounts of heat per accreted nucleon during different outbursts and may not be simply active or inactive \citep[as could still be the case after the study by][]{deibel2015strong}.

The origin of shallow heating remains unknown and thus also why its strength varies during different outbursts. One possibility \citep[as suggested by ][]{inogamov2010spread} is that the shallow heating originates from the dissipation of the accretion-generated $g$-modes in the ocean (i.e., the melted crust). Based on the $kT_\text{eff}^\infty$, we find that the ocean is deeper during outburst II than outburst III. If the shallow heat source is placed at this ocean-solid crust interface then our models indicate that more shallow heat is required during outburst III than outburst II, suggesting that the strength of heat deposition is unlikely to vary with the depth of the ocean \citep[as suggested by][]{deibel2016ocean}. 

It is interesting to note that the strength of the shallow heating seems to correlate with the outburst fluences ($Q_\mathrm{sh}$ decreases from $\sim$17 MeV nucleon$^{-1}$ to $\sim$0.3 MeV nucleon$^{-1}$ when the fluence decreases from $\sim$4.3 $\times$ 10$^{-2}$ erg cm$^{-2}$ to $\sim$4.4 $\times$ 10$^{-3}$ erg cm$^{-2}$). However, even if true for J0556, this cannot be extrapolated to other sources. Another NS LMXB XTE J1701$-$462 experienced an outburst in 2006/2007 with a similar fluence to outburst I of J0556 (\citeauthor{fridriksson2010rapid} \citeyear{fridriksson2010rapid}, \citeyear{fridriksson2011variable}; see \citetalias{homan2014strongly} for details), but J0556 needs $Q_\mathrm{sh}$ $\sim$17 MeV nucleon$^{-1}$ to explain its cooling data whereas XTE J1701$-$462 needs only $Q_\mathrm{sh}$ $\sim$0.1 MeV nucleon$^{-1}$ \citep{page2013forecasting}. Therefore, across sources there is another, unknown parameter that sets the strength of the shallow heating.

Using our \texttt{NSC\textsc{ool}} models we attempted to discern if the observed crust cooling currently follows the cooling trend defined by outburst I or if the reheating from outburst III still influences the crust temperature. We find that a model which only includes heating from outburst I, slightly undershoots our most recently obtained data point suggesting that the crust might not have yet returned to the original cooling trend. However, due to uncertainties in the data and the modelling we cannot conclusively say if this is indeed the case. Future {\it Chandra} observations will clarify this as well as probe the pasta layer present deeper in the crust.


\section*{Acknowledgements}

AP, RW, and LO are supported by a NWO Top Grant, Module 1, awarded to RW. JH acknowledges financial support from NASA grants GO6-17033X, NNX16AH25G, and NNX15AJ36G. ND is supported by an NWO Vidi grant. DP is supported by the Consejo Nacional de Ciencia y Tecnolog{\'\i}a with a CB-2014-1 grant $\#$240512. DA acknowledges support from the Royal Society. This work was benefitted from support by the National Science Foundation under Grant No. PHY-1430152 (JINA Center for the Evolution of the Elements). ML is supported by EU's Horizon 2020 programme (Marie Sklodowska-Curie Fellowship; grant nr. 702638). The authors wish to thank Mike Nowak for discussions on applying the Chandra pile-up model.

\FloatBarrier

\bibliographystyle{apj}

\end{document}